\documentclass{cpeauth}

\usepackage{t1enc,lmodern}
\usepackage{verbatim,moreverb,alltt} 
\usepackage{url}
\usepackage{graphicx}

\begin{document}
\def\cop{Copyright \copyright\ 20xx John Wiley \&\ Sons, Ltd.}

\CPE{0}{0}{00}{00}{20xx}

\runningheads{J.Ph Chancelier, B. Lapeyre and J\'er\^ome Lelong}{Using Premia and Nsp}

\title{Using Premia and Nsp for Constructing\\ a Risk Management Benchmark
  for Testing Parallel Architecture}

\author{Jean-Philippe Chancelier\corrauth\comma\footnotemark[2],
  Bernard Lapeyre\footnotemark[3] and J\'er\^ome Lelong\footnotemark[4]}


\longaddress{Universit\'e Paris-Est, CERMICS, \'Ecole des Ponts, 6 \& 8 av. B. Pascal, 77455 Marne-la-Vall\'ee, France}

\corraddr{Jean-Philippe Chancelier,  Universit\'e Paris-Est, CERMICS, \'Ecole des Ponts, 6 \& 8 av. B. Pascal, 77455 Marne-la-Vall\'ee, France}

\footnotetext[2]{E-mail: jpc@cermics.enpc.fr}
\footnotetext[3]{E-mail: bl@cermics.enpc.fr}
\footnotetext[4]{Laboratoire Jean Kuntzmann, Universit\'e de Grenoble et CNRS, BP 53,
38041 Grenoble C\'edex 9, FRANCE.\\
E-mail : jerome.lelong@imag.fr}

\cgs{This work was supported by the French Research Agency (ANR GCPMF)}

\received{}
\revised{}
\noaccepted{}

\begin{abstract}
  Financial institutions have massive computations to carry out overnight which are
  very CPU demanding. The challenge is to price many different products on a
  cluster-like architecture. We have used the Premia software to valuate the
  financial derivatives. In this work, we explain how Premia can be embedded into
  Nsp, a scientific software like Matlab, to provide a powerful tool to valuate a
  whole portfolio. Finally, we have integrated an MPI toolbox into Nsp to enable the
  use of Premia to solve a bunch of pricing problems on a cluster. This unified
  framework can then be used to test different parallel architectures.
\end{abstract}



\section{Introduction: The context of risk evaluation}

Banking legislation (Bale II\cite{BaleII}\cite{BaleIIpdf}) imposes to financial
institutions some daily evaluation of the risk they are exposed to because of
their market positions. The main investment banks own very large portfolios of
contingent claims (several thousands of claims, 5000 being a realistic
estimation).

For a given contingent claim and model parameter set, the evaluation of the
price (or other risk features such as delta, gamma, vega, \ldots) requires a
computation time which can greatly vary, from a few milliseconds (for standard
options in the Black Scholes model) to dozens of minutes (for American
options on a large number of underlying assets).

A model is specified by several parameters: volatility, interest rate, \ldots
and, in the context of risk evaluation, it is necessary to price the contingent
claims for various values of these model parameters to measure their
sensibilities to the parameters. As a consequence, a huge number of atomic
computations (around $10^6$) is necessary to evaluate the risk of the whole
portfolio. These computations must be done on a daily basis to provide an
evaluation of the position of the bank to the risk control organism. They are so
complex that financial institutions often need to use very large clusters with
up to several thousands of nodes.

Being able to have free access to both a realistic portfolio description and
some model parameters would be especially useful for benchmarking
(software/hardware) parallel architectures.  Unfortunately, in our knowledge, no
such information exists mostly because of obvious confidentiality
considerations.  Moreover, the evaluation of a complex portfolio needs a lot of
elaborate algorithms which are seldom freely available in a unified framework.

In this work, we propose a software architecture for constructing realistic
models and portfolios based on freely available softwares: Premia, MPI and Nsp.
Premia\cite{premia} will be the library used to compute financial product prices
and MPI will be used to control parallelism.  Finally, we use a software, with a
Matlab like syntax, Nsp~\cite{nsp} to provide a unified access to MPI and Premia
primitives. Using this unified framework, we are able to generate parametrised
benchmarks, to save them and to control the parallel architecture (grids,
clusters,\ldots).

We emphasise that Nsp and some implementations of MPI are available under the
GPL licence and that Premia is freely available for test and experimentation
purposes. Moreover, these softwares have been successfully compiled on the most
widely used operating systems (Windows, Linux, Mac OS X)  and their deployment
on a cluster is quite easy.  Such an environment is a step to define
standardised benchmarks useful for the evaluation and, we hope, the conception
of parallel architectures.

\section{Premia: a library for numerical computations in finance}

Premia is a research software devoted to the pricing and hedging of financial
derivatives (see \cite{lambertonlapeyre}  for an introduction), which  is a major
issue for financial institutions.  The algorithms are accompanied by a solid
scientific documentation. The software is developed in the framework of the MATHFI
research team uniting scientists working in probability and finance from INRIA and
\'Ecole des Ponts.

This project keeps track of the most recent advances in the field of computational
finance. It focuses on the implementation of numerical analysis techniques for both
probabilistic and deterministic numerical methods.  An important feature of the
Premia platform is its detailed documentation which provides extended references in
option pricing.  Besides being a single entry point for accessible overviews and
basic implementations of various numerical methods, the aim of the Premia project is
to be a powerful testing platform for comparing different numerical methods. 

Premia is developed in interaction with a consortium of financial institutions or
departments presently composed of : Calyon, Natixis, Soci\'et\'e G\'en\'erale. The
members of the consortium support the development of Premia and help to determine the
directions in which the project should evolve.

Premia is a fairly complete library with regards to what is currently used in
advanced finance. Obviously Premia cannot claim to be an exhaustive option pricer,
but it is an easy task to add any new pricing algorithms using the Premia framework. 
Once done, calling the new algorithm from Nsp does not require any additional work.

In its current public release, it contains finite difference
algorithms, tree methods and Monte Carlo methods for pricing and hedging European and
American options on equities in several models going from the standard Black-Scholes
model to more complex models such as local and stochastic volatility models and even
L\'evy models. Sophisticated algorithms based on quantisation techniques or Malliavin
calculus for European and American options are also implemented. More recently,
various interest rate and credit risk models and derivatives have been added.

\section{Tools}

\subsection{Nsp}
\label{nsp}

Nsp is a Matlab like scientific software package developed under the GPL licence.  It
is a high-level scripting language  which gives an easy access to efficient numerical
routines.  It can be used either as an  interactive computing environment or as a
programming language. It supports  imperative programming and features a dynamic
typing system and automatic memory management.  It contains an internal class system
with simple inheritance and interface implementations, this class system is visible at
the Nsp programming level but not extendable at the Nsp level.  When used as an
interactive computing environment, it comes with online help facilities  and an easy
access to GUI facilities and graphics.

A large set of libraries is available and it is moreover easy to implement new
functionalities. It requires to write some wrapper code also called interfaces to
give  glue code between the external library and Nsp internal data. The interface
mechanism  can be either static or dynamic, which makes it possible to build
toolboxes. 

Nsp shares many paradigms with other Matlab like scientific softwares as  for
example: Matlab, Octave, ScilabGtk\cite{scilabgtk}\cite{scilab} and also with
scripting  languages such as Python for instance.

Two typical toolboxes were used in this work. The first one is the Nsp Premia toolbox
which gives access at Nsp level to the Premia financial library. The second one is a
MPI interface giving access at Nsp level to mainly all MPI-2 functions.

\subsection{MPI toolbox for Nsp}
\label{sec-mpi}

Having a direct access to MPI functions within a scripting language can be
very useful for many aspects. The main advantage is that it gives an easy way
to get  familiar with the large set of MPI functions which can be tested
interactively.  It also hides the tedious work of packing and unpacking
complex data since a scripting  language contains high level data and the
packing and unpacking of such data can be hidden to  the user.

Similar toolboxes are available. As for example, MPITB\cite{mpitb} is a toolbox
initially developed by Javier Fern\'andez Baldomero and Mancia Anguita which provides
such a full MPI interface for the Matlab and Octave languages. The MPINSP toolbox
follows the  same philosophy and was implemented using the Nsp interface language.
The Matlab version of the MPItb toolbox is implemented through wrapper code which are
called  mexfiles and since a mexlib interface library is available in Nsp it could
have been possible to make the Matlab  toolbox work in Nsp with mainly no additional
work. Nevertheless, for a better efficiency and flexibility  the MPINSP toolbox has
been directly written using the Nsp interface API.

Now, we give some examples to highlight facilities that are given inside Nsp to
access MPI primitives.  It is possible to launch a master Nsp and then to
spawn Nsp slaves, this is done by using the  \verb+MPI_Comm_spawn+ primitive
as shown on Fig. \ref{Poo}:

\begin{figure*}[htbp]
\begin{verbatim}
MPI_Init();
COMM =mpicomm_create('SELF');
INFO_NULL=mpiinfo_create('NULL');
cmd = "exec(''src/loader.sce'');MPI_Init();";
cmd = cmd + "parent=MPI_Comm_get_parent();";
cmd = cmd + "[NEWORLD]=MPI_Intercomm_merge(parent,1);";
nsp_exe = getenv('SCI')+'/bin/nsp';
args=["-name","nsp-child","-e", cmd];
[children,errs]= MPI_Comm_spawn(nsp_exe,args,1,INFO_NULL,0,COMM);
// child will execute cmd
[NEWORLD] = MPI_Intercomm_merge (children, 0);
\end{verbatim}
  \caption{MPI primitives at Nsp level \label{Poo}}
\end{figure*}

The code given in Fig. \ref{Poo} will start a new Nsp which will execute
the transmitted \verb+cmd+
to start interacting with the master through a merged communicator. Note that
the interface between Nsp and MPI does not just consist in a set of functions but also
on new Nsp object devoted to MPI. For example \verb+mpicomm_create+
creates a new Nsp communicator object which internally contains a MPI communicator.
Since starting a set of Nsp slaves is a classic task, the previous code can
be written in a Nsp function \verb+NSP_spawn+ and it is then possible to
start \verb+n+ slaves by the simple Nsp command
\begin{verbatim}
NEWORLD=NSP_spawn(n);
\end{verbatim}

It is possible to transmit and receive almost all the Nsp objects using the
\verb+MPI_Send_Obj+ and \verb+MPI_Recv_Obj+ Nsp functions. These two functions
use the fact that almost all the Nsp objects can be serialized into a \verb+Serial+
object. The two functions \verb+MPI_Send_Obj+ and \verb+MPI_Recv_Obj+ use internal
serialization and packing to transparently transmit Nsp Objects.

\begin{verbatim}
-nsp->A=list('string',%t,rand(4,4));
-nsp->MPI_Send_Obj(A,rank,TAG,MCW)
-nsp->B=MPI_Recv_Obj(rank,TAG,MCW)
B       =               l (3)
 (
  (1)   =               s (1x1)

    string
  (2)   =                b (1x1)

   | T |
  (3)   =               r (4x4)

   |  0.89259  0.69284  0.10172  0.85434 |
   |  0.08482  0.67768  0.63584  0.16133 |
   |  0.25667  0.42840  0.73767  0.29179 |
   |  0.65078  0.37258  0.67447  0.23511 |
 )
\end{verbatim}

It gives us a very easy way to transmit a Premia problem to a Nsp slave.
Moreover it is easy to transmit jobs to Nsp slaves as Nsp strings.

For standard objects such as non sparse matrices, cells, lists and hash tables
it is possible to use \verb+MPI_Send+ directly or combined with the
\verb+MPI_Pack+ function.

\begin{verbatim}
A=[%t,%f];
B={'foo',[1:4],'bar',rand(100,100)};
H=hash_create(A=A,B=B);
P=MPI_Pack(H,MCW),
MPI_Send(P,randk,TAG,MCW)
\end{verbatim}

Receiving the transmitted packed data is also easy. A \verb+mpibuf+ object
can be created at Nsp level with a proper size and be given to the \verb+MPI_Recv+
function for receiving the transmitted packed data. A call to \verb+MPI_Unpack+ will
then recreate a Nsp object.

\begin{verbatim}
[stat]=MPI_Probe(-1,-1,MCW)
// size in characters
[elems]=MPI_Get_elements(stat,'')
B=mpibuf_create(elems);  // create a receive buffer
...
stat=MPI_Recv(B,randk,TAG,MCW);
H1=MPI_Unpack(B,MCW);
\end{verbatim}

Moreover, it is possible to serialize objects at Nsp level and transmit them.
Note, that in that case, \verb+MPI_Recv_Obj+ will directly unseal the
received Serial object.

\begin{verbatim}
-nsp->A=sparse(rand(2,2));
-nsp->S=serialize(A);
-nsp->MPI_Send_Obj(S,rank,TAG,MCW)
...
-nsp->B=MPI_Recv_Obj(rank,TAG,MCW);
-nsp->B.equal[A]
ans     =                b (1x1)

 | T |
\end{verbatim}

The serialization of objects is very similar to the binary format used to save
and load objects in Nsp since serialization just redirects the binary savings
of objects to a string buffer. Therefore, it is possible to save a set of
objects in a file and then directly load the file content in a serialized
object. It gives us an efficient way of transmitting Nsp data stored in a file
to MPI slaves. We illustrate in the next script the \verb+sload+ function~:

\begin{figure*}[h]
\begin{verbatim}
-nsp->H.A = rand(4,5);
-nsp->H.B = rand(4,1);
-nsp->save('/tmp/saved.bin',H);
-nsp->S=sload('/tmp/saved.bin')  // we directly create a Serial object
S	= <302-bytes>		 serial
-nsp->H1=S.unserialize[];
-nsp->H1.equal[H]
ans	=		 b (1x1)

| T |
\end{verbatim}
\caption{The {\tt sload} function}
\label{fig-sload}
\end{figure*}

We have recently introduced in Nsp the possibility to compress the serialized
buffer used in serialized objects. The \verb+unserialize+ method can then
transparently manage unserialization of compressed and non compressed Serial
objects. Using this facility to test if it can improve the MPI transmission of
Premia problems was not studied in this paper but it is left for future
developments and tests. In some Premia problems, a large set of data contained
in a file has to be embedded and transmitted with the problem, we imagine that
compressed serialization could be useful in those cases. Moreover,
compression, which takes most of the CPU time, can be done off line when
preparing a set of problems.

\begin{verbatim}
-nsp->A=1:100;
-nsp->S=serialize(A)
S	= <842-bytes>		 serial
-nsp->S1=S.compress[]
S1	= <248-bytes>		 serial
-nsp->A1=S1.unserialize[];
-nsp->A1.equal[A]
ans	=		 b (1x1)

| T |
\end{verbatim}

A large file called \verb+TUTORIAL.sce+ can be used to interactively
to learn MPI in general and also its Nsp interface. This file is a
simple Nsp adaptation of the excellent MPITB tutorial for Matlab \cite{mpitb}.

\subsection{Premia toolbox for Nsp}
\label{sec-premia-tb}

\begin{figure}
  \centering \includegraphics[width=8cm]{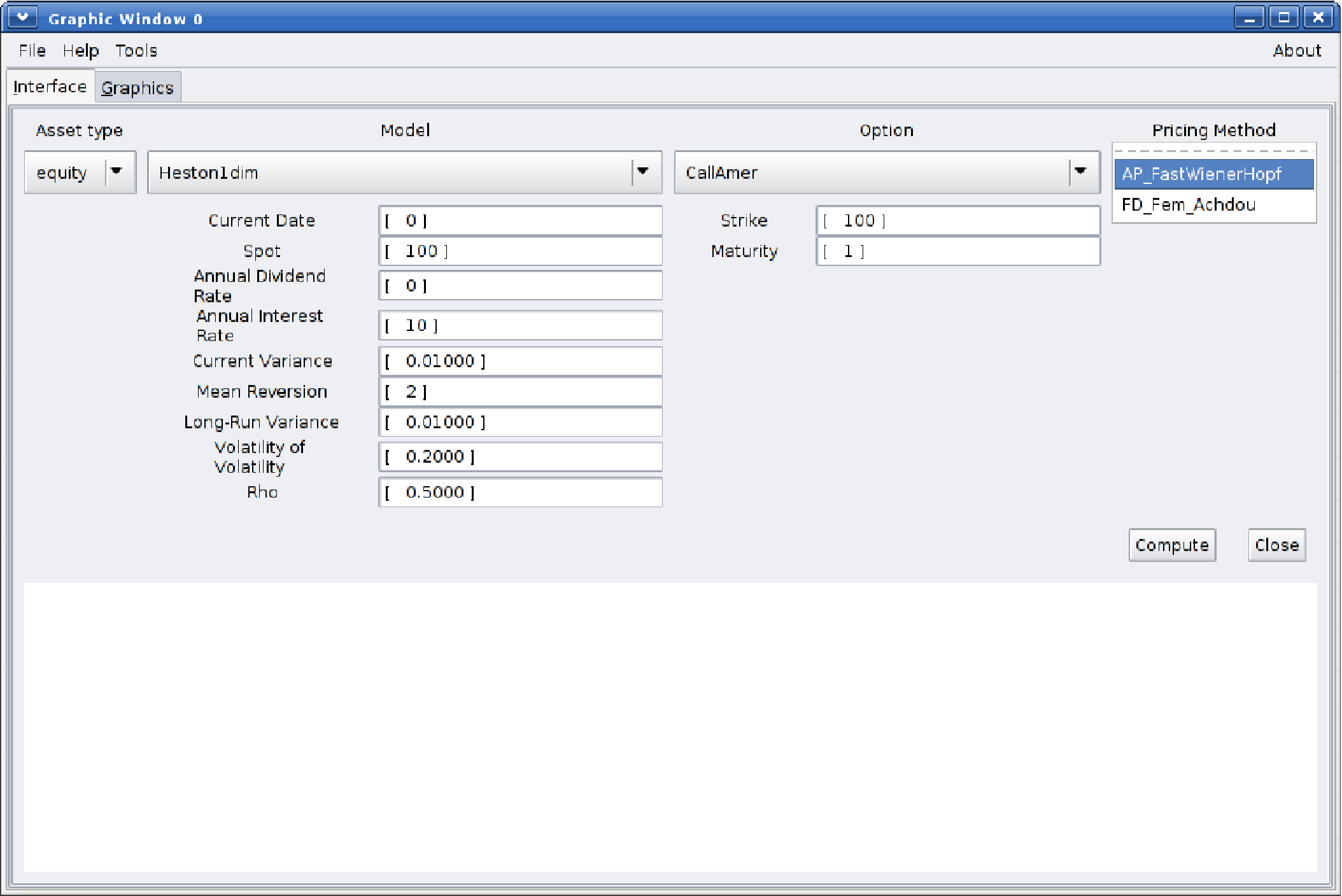}
  \caption{Premia/Nsp interface}
  \label{premia-gui}
\end{figure}

For long, the only way to use Premia was from the command line. With the
growing of Premia every year, the need of a real graphical user interface has
become more and more pressing. The idea of embedding the Premia library in a
Matlab like scientific software has come up quite naturally. Unlike a
standalone graphical user interface, embedding Premia into such a scientific
software provides two ways of accessing the library either through the
scripting language or using the graphical capabilities of the software (see
Figure~\ref{premia-gui}). The possibility of accessing the Premia functions
directly at the interpreter level makes it possible to make Premia interact
with other toolboxes. Since the licence of Premia gives right to freely
distribute the version of Premia two year older that the current release, it
was important that the scientific software used can be freely obtained and has
extensive graphical feature.  Nsp fulfilled all these conditions.

The inheritance system of Nsp enables us to easily add new objects in the
interpreter. This is how we introduced a new type named {\em PremiaModel},
through which the wide range of pricing problems described in Premia and their
corresponding pricing methods are made available within Nsp. The results
obtained in a given problem can be used in any post-treatment routines as any
other standard data.

For practitioners, the daily valuation of complex portfolios is a burning
issue to which we tried to answer using MPI/Nsp/Premia. Given a bunch of
pricing problems to be solved, which are implemented in Premia, how can we
make the most of Nsp and the two previously described toolboxes? First, we
needed a way to describe a pricing problem in a way that is understandable by
Nsp so that it can create the correct instance of the {\em PremiaModel}
class. We implemented the \verb+load+ and \verb+save+ methods for such an
instance relying on the XDR library (eXternal Data Representation). This way,
any {\em PremiaModel} object can be saved to a file in a format which is
independent of the computer architecture; these files can be reloaded later by
any Nsp process. Then, a bunch of pricing problems can  be represented by a
list of files created either from the scripting language or using the
graphical interface. Let us give an example of how to create such a file. To
save the pricing of an American call option in the one dimensional Heston
model using a finite difference method, one can use the following instructions
\begin{verbatim}
P = premia_create()
P.set_asset[str="equity"]
P.set_model[str="Heston1dim"]
P.set_option[str="PutAmer"]
P.set_method[str="MC_AM_Alfonsi_LongstaffSchwartz"]
save('fic', P)
\end{verbatim}
Creating an instance of the {\em PremiaModel} class and setting its parameters
are very intuitive.  The object saved in the file \verb+fic+ can be reloaded
using the command \verb+load('fic')+.

To solve this list of problems, we could use a single Nsp process but as the
problems are totally independent it is quite natural to try to solve them in
parallel using the MPI toolbox presented in Section~\ref{sec-mpi}. The master
process reads all the files and creates the corresponding instances of the
{\em PremiaModel} class. Then, each instance is serialized and sent to a given
remote node using MPI's communication functions.

\section{Practical experiments}
\label{sec-practical}

\begin{figure*}[t]
\begin{verbatim}
if ~MPI_Initialized() then   MPI_Init();end 
MPI_COMM_WORLD=mpicomm_create('WORLD');
[mpi_rank] = MPI_Comm_rank (MPI_COMM_WORLD);	
[mpi_size] = MPI_Comm_size (MPI_COMM_WORLD);
    
if mpi_rank <> 0				// Slave part
  while %t then 
    name = MPI_Recv_Obj(0,TAG,MPI_COMM_WORLD); // receives the name
    if name == '' then break; end 
    [stat]=MPI_Probe(-1,-1,MPI_COMM_WORLD)
    [elems]=MPI_Get_count(stat);
    pack_obj=mpibuf_create(elems);  // creates a buffer to store the packed obejct
    stat=MPI_Recv (pack_obj, 0, TAG, MPI_COMM_WORLD); // receives the packed object
    ser_obj = MPI_Unpack (pack_obj, MPI_COMM_WORLD); // unpack
    P = unserialize(ser_obj); // unserialize
    P.compute[]; L = P.get_method_results[];
    MPI_Send_Obj(L(1)(3),0,TAG,MPI_COMM_WORLD); // sends the results back
  end
else				// Master part
  Nt= size(Lpb, '*'); 
  nb_per_node = floor (Nt / (mpi_size-1));  
  slv = 1;
  for pb=Lpb(1:mpi_size-1)'			// send 
    send_premia_pb (pb, slv); slv = slv + 1;
  end
  res=list();
  Lpb(1:mpi_size-1)=[];
  for pb=Lpb'
    [sl, result] = receive_res ();
    res.add_last[list(sl, result)];
    send_premia_pb (pb, sl);
  end
  for slv=1:mpi_size-1	  // we still have mpi_size-1 receives to perform
    [sl, result] = receive_res ();
    res.add_last[list(sl, result)];
  end
  for slv=1:mpi_size-1   // tell all slaves to stop working 
    MPI_Send_Obj([''],slv,TAG,MPI_COMM_WORLD);
  end
  save('pb-res.bin',res);
end
\end{verbatim}
  \caption{A sample script for creating a parallel portfolio pricer}
  \label{load-balance}
\end{figure*}

\begin{figure}
\begin{verbatim}
// Loads a Premia object, serializes and packs it before sending it to the
// process wih number slv
function send_premia_pb( name, slv )
  load(name);
  ser_obj = serialize (P)
  MPI_Send_Obj (name,slv,TAG,MPI_COMM_WORLD); // send name
  pack_obj = MPI_Pack (ser_obj, MPI_COMM_WORLD); // pack
  MPI_Send (pack_obj,slv,TAG,MPI_COMM_WORLD);  // send the packed object
endfunction

function [sl, result] = receive_res ()
  [stat] = MPI_Probe(-1,-1,MPI_COMM_WORLD);
  sl = stat.src;
  result = MPI_Recv_Obj(sl,TAG,MPI_COMM_WORLD);
endfunction
\end{verbatim}
  \caption{Sending a Premia object}
  \label{send-premia-obj}
\end{figure}

A typical usage example of our MPI/Nsp/Premia framework is the evaluation of a
large portfolio consisting of hundreds or even thousands of options. The
pricing of a single option is not carried out using parallel computations but
instead each option is priced on a single processor and because we have many
processors at hand we can price several options simultaneously.  Although
load-balancing for parallel computation is a very active field of research, we
have restricted to a simplified ``Robbin Hood'' strategy for our tests. The
codes of Fig.~\ref{load-balance} and Fig.~\ref{send-premia-obj} describe the
load-balancing strategy used in our examples.  First, the master sends one job
to each slave and as soon as a slave finishes its computation and sends its
answer back, it is assigned a new job. This mechanism goes on until the whole
portfolio has been been treated.

We considered the examples of portfolios described in
Sections~\ref{sec-regression}, \ref{sec-comms-example},
\ref{sec-portfolio-example}. In Section~\ref{sec-premia-tb}, we explained how
a pricing problem can be saved in a file relying on the XDR library,
henceforth in our examples, a portfolio will be a collection of files, each
file describing a precise pricing problem.

In Tables~\ref{tab-tests} and~\ref{tab-comms}, the {\it Time} columns give the
computation time in seconds whereas the {\it Speedup ratio} columns give the
ratio $\frac{\text{CPU time for 1 CPUs}}{\text{n } \times \text{ CPU time for
n CPUs}}$. When this ratio becomes close to $1$, it indicates that a linear
speedup has been achieved. The columns are labelled according to the way the
{\it PremiaModel} objects are passed from the master to a slave. There are
three different labels : {\it full load}, {\it NFS}, {\it serialized load}.
The label {\it full load} means that the master reads the content of the file
describing the {\it PremiaModel} object, then creates the object, serializes
it, packs it and sends it to a slave, which in turn performs the different
operations the other way round to recreate the {\it PremiaModel} object. This
way of transmitting objects highlights that the object created by the master
would actually be useless if we could create a serialized {\it PremiaModel}
object directly from the file in which it is saved. Going directly from the
file to the serialized object without actually creating the object itself is
precisely the purpose of the \verb!sload! function (see Fig.~\ref{fig-sload}
for a description of the function). This more direct way of transmitting an
object is referred to by the {\it serialized load} label in the tables. The
cluster on which all the tests were carried out use a NFS file system, which
makes it possible for the master to only send the name of the file to be read
and let the slave read the file content instead of creating the object and
sending it to the slave. The use of the NFS file system is referred to by the
{\it NFS} label in the different tables.

All our numerical tests were carried out on a $256-$PC cluster of SUPELEC.
Each node is a dual core processor : INTEL Xeon--3075 2.66 GHz with a front
side bus at 1333Mhz. The two cores of each node share 4GB of RAM and all the
nodes are interconnected using a Gigabit Ethernet network. In none of the
experiments, did we make the most of the dual core architecture since our code
is one threaded. Hence, in our implementation a dual core processor is
actually seen as two single core processors.

\subsection{Premia's regression tests}
\label{sec-regression}

The first example we studied has been brought to our knowledge by the Premia
development team which uses a bunch of non-regression tests to make sure that
a change in the source code does not alter the behaviour of any
algorithm. These non-regression tests consist in a single instance of any
pricing problem which can be solved using Premia --- a pricing
problem corresponds to the choice of a model for the underlying asset, a
financial product and a pricing method for computing the pricing and sometimes
also the delta (first derivative of the option price with respect to the spot
price). Several sets of these tests exist with different parameters and are
run at least once a day. This motivated us to implement a parallel version of
these non-regression tests; the speedups we managed to achieved are reported
in Table~\ref{tab-tests}, which shows that for a number of nodes less than $16$
we could achieve an almost linear speedup. The pricing problems are sent using
the \verb!sload! method but changing the way of sending problems has pretty
much no effect on the computation time and speedup ratio because the
communication time is negligible compared to the computation time of a single
pricing problem. However, the decreasing of the speedup ratio when more than
$16$ nodes are used indicates that the computation time of a single problem is
too short. One way of improving the speedup ratio would be to create bunches
of several pricing problems and send them all together which would
considerably reduce the latency induced by communications: it is always
advisable to send a single large message rather several smaller messages.

\tabcap{14cm}
\begin{table}
  \caption{Speedup table for the non-regression tests of {\it Premia}.}
  \label{tab-tests}
  \begin{center}
    \begin{small}
  \begin{tabular}{c@{\hspace{0.5cm}}c@{\hspace{0.5cm}}c}
    \toprule
    number of & Time  & Speedup ratio \\
    CPUs & & \\
    \midrule
    2	& 838.004	& 1	        \\
    4	& 285.356	& 0.9789	\\
    6	& 172.146	& 0.973597	\\
    8	& 124.78	& 0.959407	\\
    10	& 97.1792	& 0.958142	\\
    16	& 67.9677	& 0.821963	\\
    32	& 45.6611	& 0.592023	\\
    64	& 34.2828	& 0.387998	\\
    96	& 31.4682	& 0.280317	\\
    128	& 30.5574	& 0.215937	\\
    160	& 16.1006	& 0.327347	\\
    192	& 30.7013	& 0.142908	\\
    224	& 30.5024	& 0.123199	\\
    256	& 31.3172	& 0.104935	\\
    \bottomrule
  \end{tabular}
  \end{small}
\end{center}
\end{table}

\subsection{A toy portfolio for discriminating communication strategies}
\label{sec-comms-example}

The purpose of this second example is to test the different ways of sending
pricing problems. For this we considered a portfolio of $10,000$ vanilla
options which can be priced using closed-form formula. A single price
computation is then very fast and the time spent in communication is easily
highlighted. From the comparison of the second and sixth columns of
Tab.~\ref{tab-comms}, it clearly appears that it is always better to use the
\verb!sload! method which consists in creating the serialized object directly
from the file containing the object rather than first creating the object
itself and then serializing it. The computation times obtained when relying on
the {\it NFS} file system for sending the pricing problems are more difficult
to analyze. Until the number of nodes used is less than $12$, the \verb!sload!
method performs better than the use of {\it NFS} but when the number of nodes
keeps on increasing the use of the {\it NFS} file system becomes faster. One
should keep in mind that the {\it NFS} file system uses a caching system which
makes the following access to the same files much faster than the first one.
This remark explains the huge difference in computation time between $2$ and
$4$ nodes in the fourth column of Table~\ref{tab-comms}. Then, the comparison
with the {\it NFS} file system may be highly biased  and {\it NFS} does not
probably outperform the \verb!sload! method so much on a clean run with a
new portfolio. The only objective comparison is between the {\it full load}
and {\it serialized load}, the latter is always the faster.

\tabcap{14cm}
\begin{table}
  \caption{Comparison of the different ways of carrying out the
    communications.}
  \label{tab-comms}
  \begin{small}
\begin{tabular}{ccccccc}
    \toprule
  number of & Time  & Speedup  ratio & Time & Speedup ration & Time  & Speedup ratio\\
  CPUs & full load & full load & NFS & NFS & serialized load & serialized load\\
  \midrule
  2  & 8.85665   & 1         & 16.3965   & 1         & 7.17891   & 1 \\
  4  & 3.55046   & 0.831503  & 4.91225   & 1.11263   & 1.73774   & 1.37706   \\
  8  & 3.86341   & 0.327492  & 2.52961   & 0.925974  & 1.81472   & 0.565132  \\
  10 & 4.06038   & 0.24236   & 2.08968   & 0.871824  & 1.87771   & 0.424802  \\
  12 & 3.9264    & 0.205061  & 1.77673   & 0.838952  & 1.88571   & 0.346091  \\
  14 & 3.9624    & 0.171937  & 1.57676   & 0.799912  & 1.81372   & 0.30447   \\
  16 & 4.05038   & 0.145775  & 1.40579   & 0.777572  & 1.9367    & 0.247118  \\
  18 & 3.9524    & 0.131813  & 1.27181   & 0.758371  & 1.9497    & 0.216591  \\
  20 & 4.13337   & 0.112775  & 1.17682   & 0.73331   & 1.87272   & 0.201759  \\
  24 & 3.77643   & 0.101967  & 1.02784   & 0.69358   & 1.84772   & 0.168925  \\
  28 & 3.9504    & 0.0830357 & 0.928859  & 0.653789  & 1.77273   & 0.149986  \\
  32 & 4.35934   & 0.0655371 & 0.848871  & 0.623086  & 1.83072   & 0.126495  \\
  36 & 4.05938   & 0.0623364 & 0.786881  & 0.595353  & 1.75773   & 0.116691  \\
  40 & 4.06538   & 0.0558604 & 0.832873  & 0.504787  & 1.81572   & 0.101378  \\
  45 & 4.12437   & 0.0488044 & 0.768884  & 0.484661  & 1.78273   & 0.0915209 \\
  50 & 4.19136   & 0.0431239 & 0.738887  & 0.452874  & 1.70474   & 0.0859417 \\
  \bottomrule
\end{tabular}
\end{small}
\end{table}

\subsection{A realistic portfolio valuation}
\label{sec-portfolio-example}

This last example comes from the risk evaluation which every financial
institution has to carry out on a daily basis. Our aim was to create a large
portfolio representative of the numbers of pricing problems and of the
computation cost. These portfolios are really challenging for parallel
computations because the time needed to compute a single price varies a lot
between the different financial derivatives composing the portfolio.

\paragraph*{Portfolio description.}

We tried to accurately reproduce the daily computation load every bank has to
face for the evaluation of its risk exposure. Even though, Premia is able to
price derivatives on many different kinds of underlying assets such as
interest rates, commodities, credits or even inflation, we have restricted to
equity derivatives for our tests. We built a portfolio of $7931$ contingent
claims on stocks.

Among the equity derivatives, the easiest to price are the so--called plain
vanilla options which are European call or put options in the Black \& Scholes
model; closed-form formulas are available for their evaluations. Our portfolio
contains $1952$ such call options with maturities quarterly distributed between
$4$ months and $8$ years and strikes uniformly varying between $70\%$ and
$130\%$ of the spot price with a step of $1\%$. 

Some options such as barrier options are more complex to price and partial
differential equation techniques are often used. Even though closed-form
formula for their prices do exist in the Black \& Scholes model, they cannot
be extended to more complex models whereas partial differential equation (PDE)
techniques are more widely applicable. We consider in our portfolio $1952$
down and out call options with maturities and strikes varying as in the
previous example. Because of the barrier clause in the option, the PDE must be
solved with a very thin time step, namely one time step every $2$ days.

So far, we have only considered one dimensional products but many of them
involve as many underlying assets as $40$ (as for the {\it Cac 40} index). These
high dimensional products are very hard to price and one has to resort
to Monte--Carlo techniques to evaluate these derivatives. We usually use $10^6$
samples for the Monte--Carlo simulations. We have included $525$ put options on
a $40$ dimensional basket with regularly distributed maturities between $0.2$
and $5$ years with a step of $0.2$ year and strikes uniformly varying between
$90\%$ and $110\%$ of the spot price with a step of $1\%$.

Practitioners sometimes feel the need of using little more sophisticated
models : the local volatility models which are very close to the Black \&
Scholes model but in which the volatility is not constant anymore but rather
depends on the current time and stock price. In these models, there are no
closed-form formula anymore and Monte-Carlo methods are used instead. We add
$1025$ call options in a local volatility model to our portfolio. The strikes
vary from $80\%$ to $120\%$ of the spot price and the maturities are regularly
distributed between $0.2$ and $5$ years with a step of $0.2$ year.

Finally, some options can be exercised at any time between the emission time
and a fixed time horizon : these options are called American options and can
only be priced using American Monte--Carlo algorithms or PDE techniques. Both
approaches are computationally very demanding. We added to our portfolio
$1952$ American put options priced using PDE with the same parameters as for
the plain vanilla options. The rest of the portfolio is composed of $7$
dimensional American put options with regularly distributed maturities between
$0.2$ and $5$ years with a step of $0.2$ year and strikes uniformly varying
between $90\%$ and $110\%$ of the spot price with a step of $1\%$. These
options are priced using American Monte--Carlo techniques.

To give an insight of the computational costs for each type of options, one
should keep in mind that the pricing of plain vanilla options is almost
instantaneous; the Monte--Carlo and PDE approaches for European options roughly
demand the same amount of computations (between $10$ and $30$ seconds); the
evaluation of American products is much longer than any other (above $60$
seconds).

\paragraph*{Experimental performances.}

The computation times needed to price the whole portfolio are fairly the same
no matter how the objects are sent by the master process, see
Tab.~\ref{tab-portfolio}. Even a naive load balancing as the one described in
Fig.~\ref{load-balance} enables to achieve very good speedup ratios; with
$256$ nodes, the speedup ratio is still better than $0.8$. However, still
increasing the number of nodes does not reduce the computation time
accordingly because the average cost of a single pricing problem is too small
compared to the time spent in communications. Therefore, many nodes are
waiting for some more work to do, which diminishes the speedup ratio.

\tabcap{13.5cm}
\begin{table}
  \caption{Comparison of the different ways of carrying out the
    communications.}
\label{tab-portfolio}
\begin{small}
\begin{tabular}{ccccccc}
    \toprule
  number of & Time  & Speedup ratio & Time & Speedup ratio & Time  & Speedup ratio \\
  CPUs & full load & full load & NFS & NFS & serialized load & serialized
  load\\
  \midrule
  2   & 5770.16   & 1         & 5799.66   & 1         & 5776.33   & 1 \\
  4   & 1980.35   & 0.971238  & 1939.46   & 0.996783  & 1925.29   & 1.00008   \\
  6   & 1154.05   & 0.999983  & 1161.25   & 0.998865  & 1157.22   & 0.998313  \\
  8   & 823.056   & 1.00152   & 828.07    & 1.00055   & 840.403   & 0.981897  \\
  10  & 641.166   & 0.999943  & 645.544   & 0.998239  & 641.096   & 1.00112   \\
  16  & 389.295   & 0.988139  & 389.097   & 0.993696  & 386.745   & 0.995716  \\
  32  & 187.441   & 0.993031  & 193.937   & 0.964676  & 189.354   & 0.984045  \\
  64  & 93.2008   & 0.982715  & 100.384   & 0.917062  & 94.7316   & 0.967868  \\
  96  & 61.5176   & 0.987335  & 69.7884   & 0.874774  & 63.1974   & 0.962119  \\
  128 & 46.7399   & 0.972068  & 54.8667   & 0.83232   & 47.6968   & 0.953585  \\
  160 & 38.4812   & 0.943068  & 41.9726   & 0.869039  & 41.1997   & 0.88178   \\
  192 & 31.5312   & 0.958107  & 35.7536   & 0.849278  & 33.5979   & 0.900132  \\
  224 & 27.2929   & 0.948056  & 31.3362   & 0.829948  & 31.5822   & 0.820171  \\
  256 & 24.4743   & 0.924566  & 28.2047   & 0.806382  & 27.8228   & 0.814163 \\
  320 & 26.1740   & 0.6911    &           &           & 26.7879   & 0.6760   \\
  384 & 20.0550   & 0.7512    &           &           & 22.5696   & 0.6682   \\
  512 & 19.7960   & 0.5704    &           &           & 20.1779   & 0.5602 \\
  \bottomrule
\end{tabular}
\end{small}
\end{table}

\section{Conclusion}

In this work, we explained how we could use Nsp with the MPI and Premia
toolboxes to address the difficult problem of parallelizing the evaluation of
a large portfolio. The use of Nsp makes the parallelization very easy as all
the code can be written in an intuitive scripting language. For our examples,
we chose a simplified {\it Robbin Hood} approach as far as load--balancing is
concerned and it already provides very good speedups. One way of improving the
speedups would be to improve the load balancing mechanism. The first idea is
to gather several pricing problems and send them all together to reduce the
communication latency. The bottleneck in the approach we used is that the
computation assigned to the first slave process is finished before the master
has already assigned the last slave a job. As we do not know how long a
computation will last, we cannot reorganize the works so that not all the
light works are assigned to the same CPUs. Nevertheless, one way of
encompassing this difficulty is to divide the nodes into sub-groups, each
group having its own master.  Then, each sub--master could apply a naive load
balancing but since it has fewer slave processes to monitor the speedups would
be better.

\end{document}